\documentclass[pra,aps,twocolumn,showpacs,superscriptaddress]{revtex4-2}

\usepackage{amsmath,amssymb,graphicx}

\usepackage{float}
\usepackage{comment}
\usepackage{bm}
\usepackage{bbold}

\usepackage[normalem]{ulem}
\usepackage{dcolumn}
\usepackage{hyperref}
\usepackage{xcolor}
\usepackage{braket}
\usepackage{mathtools}
\usepackage{textcomp}
\usepackage{inputenc}
\usepackage{orcidlink}

\begin{document}

\title{Precision Limits and Non-Gaussian Advantage in Optical Phase Estimation under Preparation Constraints}

\author{Víctor Izquierdo\,\orcidlink{0009-0000-9919-5706}}
\thanks{These authors contributed equally}
\email{Victor.Izquierdo@uab.cat}
\affiliation{Física Teòrica: Informació i Fenòmens Quàntics, Departament de Física, Universitat Autònoma de Barcelona, 08193 Bellaterra (Barcelona), Spain}
\author{Javier Lalueza-Puértolas\,\orcidlink{0009-0004-8988-6596}}
\thanks{These authors contributed equally}
\email{Javier.Lalueza@uab.cat}
\affiliation{Física Teòrica: Informació i Fenòmens Quàntics, Departament de Física, Universitat Autònoma de Barcelona, 08193 Bellaterra (Barcelona), Spain}

\date{\today}

\begin{abstract}
In optical quantum metrology, the scaling of the quantum Fisher information (QFI) with the mean photon number is a standard benchmark for relating precision to probe resources. For phase estimation, Heisenberg scaling corresponds to a quadratic growth of the QFI with the mean photon number, which Gaussian probes can approach. Interestingly, certain non-Gaussian probes can exhibit arbitrarily large QFI even at fixed mean photon number.   Here, we clarify the resource requirements underlying this behavior by introducing a constraint that separates physically admissible states from inadmissible ones: an envelope on the photon-number distribution that arises naturally from the physical limitations of state-generation methods. Building on this constraint and considering pure single-mode probes, we define a resource quantity, $N_\mathrm{RMS}$, fixed by the preparation constraints, and derive a finite bound on the QFI whose Heisenberg scaling is attainable. We identify a genuine metrological advantage of non-Gaussian states over Gaussian probes subject to the same constraints. Finally, we illustrate these results with an optical preparation model based on a universal gate set.
\end{abstract}

\maketitle
\section{Introduction}
\label{Sec: Introduction}
Quantum metrology exploits quantum resources to improve parameter-estimation precision beyond classical limits~\cite{Giovannetti2011, Degen2017}. An unknown parameter $\theta$ is encoded in a probe state $\rho_\theta$, from which it is inferred through measurements. For a locally unbiased estimator, the variance satisfies the quantum Cram\'er--Rao bound (QCRB)~\cite{Rao1945, Cramer1946, Braunstein94}
\begin{align}
    (\Delta\theta)^2\geq\frac{1}{m\mathcal{F}_\theta}.
    \label{Eq: Quantum Cramer-Rao Bound}
\end{align}
Here, $m$ is the number of independent experiment repetitions and $\mathcal{F}_\theta$ denotes the quantum Fisher information (QFI), which quantifies the response of the probe to a small variation in the unknown parameter.

Optical systems provide a prominent platform for quantum metrology~\cite{DemkowiczDobrzaski2015,Fadel2025}. A paradigmatic task is the estimation of a phase shift encoded by $e^{i\theta\hat n}$, where $\hat n$ is the photon-number operator. Taking the mean photon number $\langle\hat{n}\rangle$ as the resource, classical probes, described by mixtures of coherent states, satisfy $\mathcal F_\theta\leq 4\langle\hat n\rangle$, corresponding to shot-noise scaling. Nonclassical probes can surpass this bound; for instance squeezed-vacuum states exhibit $\mathcal F_\theta\sim\langle\hat n\rangle^2$ at large mean photon number, achieving Heisenberg scaling~\cite{Monras2006}.

While Gaussian states have been extensively studied and their QFI bounded at fixed mean photon number~\cite{Monras2006}, two natural questions arise: whether non-Gaussian probes can offer greater sensitivity and what limits apply to these more general states. The first question has attracted considerable recent interest; see, for instance,~\cite{guo2025metrological,gordillo2026quantum,Park2025}. However, the mean photon number alone imposes no finite universal bound on the QFI: non-Gaussian probes can exhibit arbitrarily large sensitivity even at fixed mean photon number~\cite{Rivas2012,Luis2017,Lee2019}. As previously emphasized, this resource measure can obscure the resources associated with rare, highly excited components and therefore provide an incomplete account of metrological resources~\cite{Branford2021}.

In this work, we propose a framework that resolves this issue by showing that the limitations of experimental setups impose a constraint on the photon-number distributions of admissible probes. This constraint reveals that states exhibiting divergent behavior are physically inaccessible. Within this framework, we introduce a counterpart of the mean photon number to quantify the required resources, which we call the root-mean-square photon budget, $N_{\rm RMS}$. This quantity incorporates the physical limitations of the setup and allows us to derive a finite bound on the QFI that recovers a Heisenberg limit in terms of this budget. We further compare Gaussian and non-Gaussian probes, showing that preparation constraints can
make the latter necessary to achieve Heisenberg scaling. In particular, for families of exponentially decaying envelopes with fixed decay length, the Gaussian QFI grows at most linearly with $N_{\rm RMS}$, whereas non-Gaussian probes can attain quadratic scaling. Finally, we illustrate the framework with a setup built from a universal gate set, in which the limiting resources that determine both the envelope and the photon budget of the experiment are readily identified.
\section{Preliminaries}
\label{Sec: Preliminaries}
We consider local phase estimation with pure single-mode optical probes, assuming lossless encoding and an ideal external phase reference whose resources are not counted~\cite{12Jarzyna}. We quantify sensitivity through the QFI and assume asymptotic attainability of the QCRB as the number of measurements, $m$, grows; prior uncertainty and precision at fixed total resources are beyond this scope. An unknown phase shift $\theta$ is encoded into the probe $\ket{\psi_0}$ as,
\begin{align}
    \ket{\psi_\theta}=e^{i\theta\hat{n}}\ket{\psi_0}.
    \label{Eq: Encoded state}
\end{align}
For this encoding, the QFI can be written in the following simple form~\cite{Braunstein94}
\begin{align}
    \mathcal{F}_\theta=4\mathrm{Var}_{\ket{\psi_0}}[\hat{n}].
    \label{Eq: QFI from generator n}
\end{align}
A well-studied class of probes are Gaussian states~\cite{Caves1981, Monras2006}, which QFI is upper-bounded as,
\begin{align}
    \mathcal{F}_\theta\leq8(\langle\hat{n}\rangle_{\psi_0}^2+\langle\hat{n}\rangle_{\psi_0}).
    \label{Eq: SGL}
\end{align}
This bound is attained for squeezed-vacuum states, i.e., they are optimal within this family. This bound defines the \textit{standard Gaussian limit} (SGL), which remains finite whenever the mean photon number $\langle\hat{n}\rangle_{\psi_0}$ is finite.

Therefore, when the mean photon number is the sole resource, optimal Gaussian states reach the Heisenberg limit. For Gaussian probes, fixing this first moment guarantees a finite QFI, because constraining the mean photon number bounds all higher-order moments.

\subsection{Heavy-tailed states}
\label{Subec: Heavy-tails}
Outside the set of Gaussian states, we can always construct probes with a finite mean photon number but a divergent second moment and, hence, a divergent QFI. The simplest such family consists of what we call \textit{heavy-tailed states},
\begin{align}
    \ket{\psi_\gamma} = \mathcal{N}\sum_{n=1}^\infty\frac{1}{n^\gamma}\ket{n},
    \qquad
    \gamma \in \left(1,\frac{3}{2}\right].
    \label{Eq: Heavy-Tailed distributions}
\end{align}
In Appendix~\ref{Appx: Heavy-tailed states}, we show that their QFI diverges while their mean photon number remains finite.

Although these states are normalizable and hence belong to the Hilbert space, their physical realization is problematic because their energy variance diverges. 
Their unbounded support alone is not responsible for the divergence, as coherent states also have unbounded support but finite variance. Instead, this behavior arises from the slow algebraic decay of their photon-number distribution.  

\subsection{Asymmetrically superposed states}
\label{Subec: epsilon-family}

Along these lines, Refs.~\cite{Park2025, Rivas2012} proposed probes based on asymmetric states. They superpose the vacuum $\ket{0}$ with a highly excited state $\ket{\phi}$ (with $\langle\phi|\hat n|\phi\rangle>0$ and $\braket{0|\phi}=0$) to build states with large photon-number variances. To keep the mean photon number low, the excitation is weighted with a probability amplitude $0<\epsilon\leq1$,
\begin{align}
    \ket{\psi_\epsilon} = \sqrt{1-\epsilon^2}\ket{0}+\epsilon\ket{\phi}.
    \label{Eq: Asymmetric Superposition}
\end{align}
For excitations that overlap the vacuum, the state requires an $\epsilon$-dependent normalization, which modifies the expressions below, although the physical conclusions are not altered. These states have finite energy fluctuations whenever the excitation has a finite second moment, $\langle\phi|\hat n^2|\phi\rangle<\infty$.

Since $\langle\hat{n}\rangle_{\psi_\epsilon}=\epsilon^2\langle\hat{n}\rangle_\phi$, the QFI of this state is
\begin{align}
    \mathcal{F}_\theta &=4\epsilon^2(\langle\hat{n}^2\rangle_\phi-\epsilon^2\langle\hat{n}\rangle^2_\phi) 
    \nonumber \\
    &= 4\langle\hat{n}\rangle_{\psi_\epsilon}\left(\frac{\langle\hat{n}^2\rangle_\phi}{\langle\hat{n}\rangle_\phi} - \langle\hat{n}\rangle_{\psi_\epsilon}\right).
    \label{Eq: QFI assymetric superposition}
\end{align}

Higher excitations can therefore enhance the QFI at fixed mean photon number. For instance, the Fock excitation $\ket{\phi}=\ket{N}$ with $N=\nu/\epsilon^2$, along values of $\epsilon$ for which $N$ is an integer, keeps the probe mean fixed at $\langle\hat n\rangle_{\psi_\epsilon}=\nu$ and gives
\begin{align}
    \mathcal{F}_\theta=4\nu^2\left(\epsilon^{-2}-1\right),
\end{align}
which diverges as $\epsilon\to0$ and surpasses the Heisenberg limit when the mean photon number is taken as the resource. Other excitations produce the same divergence with family-dependent prefactors.

The divergence in the preceding example requires increasingly large Fock excitations as $\epsilon\to0$. For a fixed excitation $\ket{\phi}$ with finite second moment, the QFI instead vanishes in this limit. Its maximum over the weight $\epsilon$ occurs at
\begin{align}
    \epsilon_*^2
    = \min\left\{1,
    \frac{\langle\hat n^2\rangle_\phi}
    {2\langle\hat n\rangle_\phi^2}
    \right\}.
\end{align}
Since $\langle\hat n^2\rangle_\phi\geq \langle\hat n\rangle_\phi^2$, the optimal weight satisfies $1/2\leq\epsilon_*^2\leq1$, with $\epsilon_*^2=1/2$ for a Fock excitation (see Appendix.\ref{Appx: Resources}). Fig.~\ref{fig:placeholder} illustrates analogous behavior for coherent and squeezed-vacuum excitations, with the maximum QFI attained away from $\epsilon\to0$. Thus, reducing the excitation weight does not indefinitely improve sensitivity when the excitation itself is fixed. The probe's mean photon number alone does not capture this distinction, since a small weight can compensate for an increasingly energetic excitation.

\begin{figure}  
\centering\includegraphics[width=.8\linewidth]{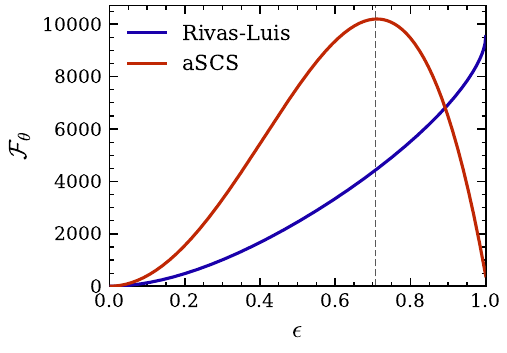}
    \caption{QFI versus $\epsilon$ for $\ket{\psi_\mathrm{RL}}= \mathcal{N}(\sqrt{1-\epsilon^2}\ket{0}+\epsilon\ket{\chi})$, with
    $\ket{\chi}=\hat S(r)\ket{0}$ at $21.4$~dB [$r=21.4\ln10/20$, zero relative phase] (blue), and for the asymmetric superposition of coherent states $\ket{\psi_\mathrm{aSCS}}=\mathcal{N}(\sqrt{ \epsilon^2}\ket{0}+\epsilon\ket{\alpha})$, with $\alpha=10$ (red). The latter peaks at $\epsilon\simeq1/\sqrt{2}$ (displaced cat), while the former is maximal for the pure squeezed vacuum ($\epsilon=1$).}
    \label{fig:placeholder}
\end{figure}

\section{Preparation constraints through photon-number envelopes}
The behavior of the states analyzed in the previous sections exposes the tension inherent in using the mean photon number as the figure of merit. This ambiguity disappears once the finite resources available for state preparation are taken into account, such as finite pumping~\cite{Gerry_Knight_2004,Scully_Zubairy_1997,16Harder,24Chinni,70Walls} or bounded squeezing~\cite{16Vahlbruch}. 
In particular, the expected photon number of a state cannot exceed the number of photons the source is able to generate. Likewise, the probability of populating Fock states above the mean photon number produced by the source cannot decay more slowly than the source's own photon-number distribution. Depending on the platform, this envelope can take the form of a hard cutoff, as in Law-Eberly-type protocols~\cite{96Law,09Hofheinz}, or of an exponential (or faster) decay, as for coherent photon sources~\cite{Gerry_Knight_2004,Scully_Zubairy_1997,16Harder,24Chinni,70Walls} or heralding protocols~\cite{16Harder}.

To unify and model these constraints, we introduce a general treatment based on a point-wise envelope, a non-negative sequence $(g_n)_n$ that bounds the photon-number distribution $q_n\coloneq |\braket{n|\psi}|^2$. Notably, this approach provides a coarse description of preparation restrictions without specifying the photon source. For a given threshold $N_\mathrm{max}$, we require the envelope to satisfy $g_n=1$ for $1\leq n\leq N_\mathrm{max}$, $0\leq g_n\leq 1$ for $n>N_\mathrm{max}$, and $\sum_{n\geq N_\mathrm{max}}n^2g_n<\infty$. Then, the admissible distributions are
\begin{align}
\label{app_admissible_set}
    \mathcal{Q}_g= \left\{q_n\geq0:\hspace{0.1cm}\sum_{n\geq0}q_n=1,\hspace{0.1cm}q_n\leq g_n\hspace{0.1cm}(n\geq N_\mathrm{max})\right\}.
\end{align}
Thus, the range $0\leq n\leq N_\mathrm{max}$ forms a free window, including an unconstrained vacuum population, while a nonzero tail allows occupation above $N_\mathrm{max}$. This constraint defines the class considered here, although additional restrictions imposed by a specific device may narrow it.

In Sec.~\ref{Sec: Example}, we illustrate how the framework translates apparatus-level restrictions into a common constraint on the probes used for phase estimation. We consider a preparation model built from a universal optical gate set, where bounds on the available gate resources constrain the output photon-number distributions and thereby determine an envelope.

\section{Limits on the precision}
Since $\mathcal{F}_\theta=4\mathrm{Var}[\hat n]$, Eq.~\eqref{Eq: QFI from generator n}, the second moment controls the sensitivity; accounting for both typical photon numbers and rare high-energy events. To compare probes under the same preparation constraints, we consider the largest second moment allowed by the envelope rather than the moment of a particular probe. Taking its square root gives a photon-number scale fixed by these constraints, which we define as the envelope's root-mean-square photon budget,
\begin{equation}\label{def_rms_resource}
    N_\mathrm{RMS}(g)\coloneq \left(\max_{q\in\mathcal{Q}_g}\mathbb{E}_q[\hat{n}^2]\right)^{1/2}.
\end{equation}
Here, $\mathbb{E}_q[\hat{n}]\coloneq\sum_{n\geq0}q_n n^2$ denotes the expectation with respect to the photon-number distribution $q$. In Appendix~\ref{app_Existence_uniqueness_RMS}, we construct the unique maximizing distribution, which need not maximize the QFI. 

For every admissible probe we derive the following upper-bound,
\begin{equation}\label{bound_QFI}
    \mathcal{F}_\theta\leq4\langle\hat{n}^2\rangle\leq 4N_\mathrm{RMS}^2(g).
\end{equation}
Thus, we obtain a Heisenberg limit in terms of the root-mean-square photon budget, fixed by the preparation constraints independently of the chosen probe. The condition $\sum_{n\geq N_\mathrm{max}}n^2g_n<\infty$ ensures that this bound is finite and holds uniformly over all admissible probes. In particular, rare high-photon-number components cannot produce unbounded sensitivity for a fixed envelope budget.

The quadratic scaling with $N_\mathrm{RMS}(g)$ is attainable up to a constant factor by states whose photon-number distributions belong to $\mathcal{Q}_g$. Denoting by $t$ the distribution that maximizes Eq.~\eqref{def_rms_resource}, we construct the pure probe
\begin{equation}\label{optimal_NG_state}
    \ket{\psi_t}=\sqrt{\frac{1+t_0}{2}}\ket{0}+\sum_{n\geq 1}\sqrt{\frac{t_n}{2}}e^{i\phi_n}\ket{n}.
\end{equation}
This state is admissible for arbitrary phases and satisfies
\begin{equation}\label{QFI_lower_bound_best_NG}
    \mathcal{F}_\theta(\psi_t)=2N_\mathrm{RMS}^2(g)-\left(\sum_{n\geq0}nt_n\right)^2\geq N_\mathrm{RMS}^2(g),
\end{equation}
where the inequality follows from $\mathbb{E}_t[\hat{n}]^2\leq\mathbb{E}_t[\hat{n}^2]=N_{\mathrm{RMS}}^2(g)$.

To define scaling, consider a specified family of envelopes $g^{(\lambda)}$, where $\lambda$ is a tunable parameter of the preparation apparatus, for example, $\lambda=N_\mathrm{max}$ at fixed tail shape and decay parameters. For each $\lambda$, we select an admissible probe and keep both the envelope and the state fixed throughout the $m$ repetitions. Scaling compares these experiments as $N_\mathrm{RMS}(g^{(\lambda)})$ increases.

For a hard cutoff, $g_n=0$ for $n>N_{\max}$, the RMS budget reduces to $N_{\mathrm{RMS}}=N_{\max}$, and Eq.~\eqref{bound_QFI} yields $\mathcal{F}_\theta\le4N_{\max}^2$. A sharper bound follows from the bounded photon-number support. By Popoviciu's inequality~\cite{Popoviciu1935}, $\operatorname{Var}(\hat n)\le N_{\max}^2/4$, hence $\mathcal{F}_\theta\le N_{\max}^2$. This bound is attained for the balanced superposition $(|0\rangle+|N_{\max}\rangle)/\sqrt2$~\cite{Lee2019}.

We now return to the state families of Sec.~\ref{Sec: Preliminaries}. The envelope constraint excludes heavy-tailed probes with divergent second moments, since we require $\sum_{n\ge1}n^2g_n<\infty$. For the asymmetrically superposed states, fixing a common envelope makes $N_{\mathrm{RMS}}(g)$ independent of the vacuum weight, even though the probe's actual photon-number moments vary with $\epsilon$. Increasing the excitation while keeping the probe's mean photon number fixed cannot produce a divergent QFI under this envelope. Indeed,
\begin{equation}
    \langle\hat n\rangle_{\psi_\epsilon}
    \frac{\langle\hat n^2\rangle_\phi}
         {\langle\hat n\rangle_\phi}
    = \langle\hat n^2\rangle_{\psi_\epsilon}
    \leq N_{\mathrm{RMS}}^2(g),
\end{equation}
so the QFI in Eq.~\eqref{Eq: QFI assymetric superposition} remains bounded. Thus, the fixed-mean divergence discussed in Sec.~\ref{Sec: Preliminaries} requires an unbounded increase in $N_{\mathrm{RMS}}(g)$, revealing a growing resource requirement that the mean photon number alone does not capture.

\section{Non-Gaussian advantage}
\label{Secc_NG_advg}
We now use this framework to characterize non-Gaussian advantage. Within a fixed envelope, every pure Gaussian state $\ket{\psi_\mathrm{G}}$ satisfies $\langle \hat{n}\rangle_{\ket{\psi_\mathrm{G}}} \leq \bar{N}(g)$, where $\bar{N}(g)=\max_{q\in\mathcal{Q}_g}\mathbb{E}_q[\hat{n}]$ is attained by the distribution $t$ that maximizes the RMS photon budget (see Appendix~\ref{app_Existence_uniqueness_RMS}). Since $x(x+1)$ is increasing for $x\geq 0$, Eq.~\eqref{Eq: SGL} yields
\begin{equation}\label{QFI_gaussian_envelope_framework}
    \mathcal{F}_\theta(\psi_\mathrm{G})\leq 8\bar{N}(g)(\bar{N}(g)+1).
\end{equation}
Note that $\bar{N}(g)$ is an auxiliary quantity fixed by the envelope, not an additional resource.

Combining Eqs.~\eqref{QFI_lower_bound_best_NG} and~\eqref{QFI_gaussian_envelope_framework}, we find
\begin{equation}\label{bound_NG_resource}
    \frac{\mathcal{F}_\theta(\psi_t)}{\mathcal{F}_\theta(\psi_\mathrm{G})}\geq \frac{N_\mathrm{RMS}^2(g)}{8\bar{N}(g)(\bar{N}(g)+1)}.
\end{equation}
Since the right-hand side depends only on the envelope $g$, this inequality provides a sufficient criterion for non-Gaussian advantage directly from the envelope constraint. Whenever it exceeds one, the probe $\ket{\psi_t}$ of Eq.~\eqref{optimal_NG_state}, which attains optimal scaling, outperforms every admissible Gaussian state. Along a family $g^{(\lambda)}$ with increasing $N_\mathrm{RMS}(g^{(\lambda)})$, the condition $\bar{N}(g)/N_\mathrm{RMS}(g)\to0$ prevents Gaussian probes from achieving Heisenberg scaling in the resource, while $\ket{\psi_t}$ retains it. Physically, the envelope allows large second moments without a comparable increase in the first moment, which limits Gaussian sensitivity.

As an example, consider the algebraically decaying envelope for $n>N_{\max}$,
\begin{equation}\label{algebraic_env_ex}
g_n=(N_{\max}/n)^p,
\end{equation}
with fixed $p>3$, so that the envelope has a finite second moment ($\sum_{n\geq N_\mathrm{max}} n^2g_n<\infty$). Increasing $N_{\max}$ enlarges the free window while preserving the envelope profile as a function of $n/N_{\max}$. Both $\bar N(g)$ and $N_{\mathrm{RMS}}(g)$ grow proportionally to $N_{\max}^{p/(p-1)}$, as we show in Appendix~\ref{Appx:algebraic_ex}, and for sufficienlty large $N_\mathrm{max}$ the right-hand side of Eq.~\eqref{bound_NG_resource} satisfies
\begin{equation}\label{alg_ex_limit_NG_adv}
\frac{N_{\mathrm{RMS}}^2(g)}{8\bar N(g)[\bar N(g)+1]}\simeq\frac{(p-2)^2}{8(p-1)(p-3)}.
\end{equation}
This limit exceeds one for $3<p<2+\sqrt{8/7}$, certifying that $\ket{\psi_t}$ outperforms every admissible Gaussian probe for sufficiently large $N_{\max}$. Thus, Eq.~\eqref{bound_NG_resource} establishes a constant-factor advantage at fixed tail exponent, although it does not yield a scaling separation in $N_{\mathrm{RMS}}(g)$.

A second criterion can be derived when the envelope decays at least exponentially at large photon numbers. We define its asymptotic exponential rate and corresponding tail length by
\begin{equation}\label{defs_tail_decay}
    \rho_g=\limsup_{n\to\infty}g_n^{1/n},\hspace{0.5cm}\ell_g=-1/\log\rho_g,
\end{equation}
with $0<\rho_g<1$. The condition $\rho_g<1$ means that an exponential upper bound holds sufficiently far in the envelope tail. Algebraic tails, and subexponential tails more generally, have $\rho_g=1$ and are not covered by this second condition.

For any admissible Gaussian probe, the squeezing parameter $r$ satisfies $\tanh{r}\leq\rho_g$, see Appendix~\ref{Appx_sec_bound_tail_decay}, independently of displacement. The envelope thus limits squeezing even when it allows $N_\mathrm{RMS}(g)$ to grow. As we further show in Appendix~\ref{Appx_sec_bound_tail_decay}, this restriction yields an upper bound on the QFI of Gaussian states in terms of $\ell_g$ and $N_\mathrm{RMS}(g)$,
\begin{equation}\label{bound_gaussian_qfi_tail_decay}
    \mathcal{F}_\theta(\psi_\mathrm{G})\leq 2\ell_g^2+4(1+2\ell_g)N_\mathrm{RMS}(g).
\end{equation}
In this bound, the first term comes from squeezing. It depends only on how fast the tail of the envelope decays, and not on $N_{\mathrm{RMS}}(g)$. The second term comes from displacement and grows linearly with $N_{\mathrm{RMS}}(g)$. Thus, increasing the budget at fixed tail decay, for example by enlarging the free window, allows at most linear growth of the Gaussian QFI with $N_{\mathrm{RMS}}(g)$.

Combining this with Eq.~\eqref{QFI_lower_bound_best_NG} gives
\begin{equation}\label{bound_NG_decay}
     \frac{\mathcal{F}_\theta(\psi_t)}{\mathcal{F}_\theta(\psi_\mathrm{G})}\geq \frac{N_\mathrm{RMS}^2(g)}{2\ell_g^2+4(1+2\ell_g)N_\mathrm{RMS}(g)}.
\end{equation}
Thus, at bounded $\ell_g$, the QFI of Gaussian probes is limited to $\mathcal{O}(N_{\mathrm{RMS}})$, whereas $\ket{\psi_t}$ achieves $\mathcal{O}(N_{\mathrm{RMS}}^2)$. More generally, $\ell_g/N_{\mathrm{RMS}}(g)\to0$ excludes Gaussian Heisenberg scaling. The advantage arises because the envelope permits increasing photon-number fluctuations while restricting Gaussian squeezing.

To illustrate the distinction between the two criteria, consider the exponentially decaying envelope for $n>N_{\max}$,
\begin{equation}\label{exp_env_ex}
g_n=2^{-(n-N_\mathrm{max})}.
\end{equation}
As we show in Appendix~\ref{Appx:exponential_ex}, the distribution $t_{N_{\max}+j}=2^{-j}$ gives: 
$\bar N(g)=N_{\max}+2$,$N_{\mathrm{RMS}}^2(g)=N_{\max}^2+4N_{\max}+6$. As $N_{\max}\to\infty$, the right-hand side of Eq.~\eqref{bound_NG_resource} approaches $1/8$ and therefore does not certify an advantage. 

Instead, computing $\rho_g=1/2$ and $\ell_g=1/\ln2$, Eq.~\eqref{bound_NG_decay} certifies an advantage growing at least linearly with $N_{\mathrm{RMS}}(g)$. Indeed, as we detail in Appendix~\ref{Appx:exponential_ex}, the best admissible Gaussian probe has $\mathcal F_\theta(\psi_{\mathrm G})\sim N_{\mathrm{RMS}}$, whereas $\mathcal F_\theta(\psi_t)\sim N_{\mathrm{RMS}}^2$. The distinction between the two criteria arises because Eq.~\eqref{bound_NG_resource} does not capture the envelope's restriction on squeezing.

Both bounds, Eqs.~\eqref{bound_NG_resource} and~\eqref{bound_NG_decay}, compare probes under the same envelope and provide sufficient conditions for non-Gaussian advantage. A lower bound of at most one therefore does not rule out an advantage. These results show that preparation constraints can impose a fundamental separation between Gaussian and non-Gaussian metrology. In particular, at fixed exponential tail decay, Gaussian probes are limited to shot-noise-like scaling in $N_{\mathrm{RMS}}(g)$, while admissible non-Gaussian probes attain Heisenberg scaling under the same constraints.

\section{Example: Universal optical gate set}
\label{Sec: Example}
We now apply the previous results to a concrete preparation model using the universal gate set $\mathcal{G}\equiv\{\hat D,\hat S,\hat R,\hat K\}$ of displacements, squeezing, rotations, and Kerr gates~\cite{99Lloyd,06Wu} (see Appendix~\ref{Appx: Details Example} for the proofs of the section). Circuit resources are quantified by the total squeezing $r_\mathrm{tot}=\sum_i|r_i|$ and displacement $|\alpha_\mathrm{tot}|=\sum_j|\alpha_j|$, which determine the envelope for Gaussian circuits. Rotations and Kerr gates preserve photon-number probabilities but alter phases relevant to subsequent squeezing and displacement. We show that the same envelope remains valid when these gates are interleaved in circuits without squeezing. For general circuits, we assume this property based on a heuristic argument in Appendix~\ref{Appx: Details Example}. Table~\ref{tab:scalings} includes only cases for which we show that the envelope remains valid.

Firstly, we compute the decay rate of the tail fixed by the resources, so that, for all $n>N_\mathrm{max}$, the admissible probes satisfy:
\begin{align}\label{tail_decay_implementation}
    q_n \leq \varepsilon(1-e^{-s_*})\,e^{-s_*\,(n-N_\mathrm{max}-1)},
\end{align}
where $0<s_*<\kappa=\ln\coth r_\mathrm{tot}$. The rate $s_*$ must be strictly smaller than $\kappa$, and it is found by minimizing $N_{\rm RMS}$. The threshold, in contrast, depends explicitly on both the total displacement and squeezing applied. Requiring Eq.~\eqref{tail_decay_implementation} for every probe gives:
\begin{align}\label{Nmax_certified}
    N_\mathrm{max}\sim\; e^{2r_\mathrm{tot}}|\alpha_\mathrm{tot}|^2.
\end{align}
Evaluating Eq.~\eqref{def_rms_resource} for this envelope, we obtain the root-mean-square photon budget:
\begin{align}\label{NRMS_example}
    N_\mathrm{RMS}^2=N_\mathrm{max}^2+\varepsilon\left[\frac{2N_\mathrm{max}}{1-e^{-s_*}}+\frac{1+e^{-s_*}}{(1-e^{-s_*})^2}\right].
\end{align}
Hence, the resources that the setup can supply, $|\alpha_\mathrm{tot}|$ and $r_\mathrm{tot}$, bound the achievable QFI through $N_\mathrm{RMS}$. Table~\ref{tab:scalings} compares representative admissible probes at equal $N_{\mathrm{RMS}}(g)$. The non-Gaussian probes scale quadratically, whereas the prefactor of the squeezed vacuum, $2/(\ln\varepsilon^{-1}+2r_\mathrm{tot})^2$, decreases with both the tolerance and the squeezing.
\begin{table}[ht!]
    \centering
    \begin{tabular}{lcc}
        \hline\hline
        Probe & State & $\mathcal{F}_\theta$ \\
        \hline
        Squeezed vacuum & $\hat S(r)\ket{0}$
        & $\simeq 2N_\mathrm{RMS}^2/(\ln\varepsilon^{-1}+2r_\mathrm{tot})^2$ \\
        Displaced cat & $\frac{\ket{0}+\ket{\alpha}}{\sqrt{2(1+e^{-|\alpha|^2/2})}}$
            & $\simeq N_\mathrm{RMS}^2$ \\
        ON & $\big(\ket{0}+\ket{N_\mathrm{max}}\big)/\sqrt{2}$
            & $N_\mathrm{max}^2\sim N_\mathrm{RMS}^2$ \\
        \hline\hline
    \end{tabular}
    \caption{Quantum Fisher information of representative admissible probes. The comparison is therefore at equal resource $N_\mathrm{RMS}(g)$. All rows use $s=s_*$ at fixed tolerance $\varepsilon$, and $\simeq$ denotes asymptotic equivalence along each path. $\varepsilon$ refers to the quantity of probability density above $N_{\rm max}$.
    }
    \label{tab:scalings}
\end{table}

\section{Discussion}
\label{Sec: Discussion}
In this work, we incorporated preparation constraints into local optical phase estimation through upper envelopes on the probe’s photon-number probabilities. We defined a finite root-mean-square photon budget fixed by these constraints and derived a Heisenberg limit in this budget whose scaling is attainable. Divergent QFI at fixed mean photon number therefore requires an unbounded preparation budget. We also established sufficient conditions for non-Gaussian probes to outperform every admissible Gaussian state under the same envelope. In particular, for exponentially decaying envelopes with fixed decay length, Gaussian QFI grows at most linearly with the budget, whereas admissible non-Gaussian probes can attain quadratic scaling. Preparation constraints can thus make non-Gaussianity necessary to achieve Heisenberg scaling.

Using an optical preparation model, we illustrated how experimentally relevant resources can be translated into photon-number envelopes. This connects our framework to the capabilities of optical setups and provides sensitivity benchmarks in terms of available preparation resources. A natural next step is to identify preparation and measurement protocols that realize the predicted advantage in concrete devices and assess its robustness to photon loss. 
\section*{Acknowledgments}
The authors thank Martí Perarnau-Llobet, John Calsamiglia and Ricard Ravell Rodríguez for fruitful discussions, advice and feedback on the manuscript. This research was supported by the ATRAE Program (Grant ATR2024154621), funded by MICIU/AEI/10.13039/501100011033.

The authors acknowledge the use of generative AI tools (Claude and ChatGPT) for the purposes of literature search, language editing of the manuscript and assistance in deriving equations. All AI-assisted content was reviewed, verified, and revised by the authors, who take full responsibility for the content of this work.
\bibliographystyle{apsrev4-2}
\bibliography{bibliography}

@article{Giovannetti2011,
  author        = {Giovannetti, Vittorio and Lloyd, Seth and Maccone, Lorenzo},
  title         = {Advances in quantum metrology},
  journal       = {Nat. Photonics},
  volume        = {5},
  pages         = {222--229},
  year          = {2011},
  doi           = {10.1038/nphoton.2011.35},
}

@article{Degen2017,
  author        = {Degen, C. L. and Reinhard, F. and Cappellaro, P.},
  title         = {Quantum sensing},
  journal       = {Rev. Mod. Phys.},
  volume        = {89},
  pages         = {035002},
  year          = {2017},
  doi           = {10.1103/RevModPhys.89.035002},
}

@article{Rao1945,
  author        = {Rao, C. R.},
  title         = {Information and the accuracy attainable in the estimation of statistical parameters},
  journal       = {Bull. Calcutta Math. Soc.},
  volume        = {37},
  pages         = {81--91},
  year          = {1945},
}

@book{Cramer1946,
  author        = {Harald Cram{\'e}r},
  title         = {Mathematical Methods of Statistics},
  publisher     = {Princeton University Press},
  year          = {1946},
  address       = {Princeton, NJ},
}

@incollection{DemkowiczDobrzaski2015,
  author        = {Demkowicz-Dobrza{\'n}ski, Rafa{\l} and Jarzyna, Marcin and Ko{\l}ody{\'n}ski, Jan},
  title         = {Quantum limits in optical interferometry},
  booktitle     = {Progress in Optics},
  editor        = {Wolf, Emil},
  volume        = {60},
  publisher     = {Elsevier},
  address       = {Amsterdam},
  pages         = {345--435},
  year          = {2015},
  doi           = {10.1016/bs.po.2015.02.003},
}

@article{Fadel2025,
  author        = {Fadel, Matteo and Roux, Noah and Gessner, Manuel},
  title         = {Quantum metrology with a continuous-variable system},
  journal       = {Rep. Prog. Phys.},
  volume        = {88},
  pages         = {106001},
  year          = {2025},
  doi           = {10.1088/1361-6633/ae00d8},
}

@book{Gerry_Knight_2004,
  author        = {Gerry, Christopher and Knight, Peter},
  title         = {Introductory Quantum Optics},
  publisher     = {Cambridge University Press},
  year          = {2004},
  address       = {Cambridge},
  doi           = {10.1017/CBO9780511791239},
}

@book{Scully_Zubairy_1997,
  author        = {Scully, Marlan O. and Zubairy, M. Suhail},
  title         = {Quantum Optics},
  publisher     = {Cambridge University Press},
  year          = {1997},
  address       = {Cambridge},
  doi           = {10.1017/CBO9780511813993},
}

@article{Caves1981,
  author        = {Caves, Carlton M.},
  title         = {Quantum-mechanical noise in an interferometer},
  journal       = {Phys. Rev. D},
  volume        = {23},
  pages         = {1693--1708},
  year          = {1981},
  doi           = {10.1103/PhysRevD.23.1693},
}

@article{Monras2006,
  author        = {Monras, Alex},
  title         = {Optimal phase measurements with pure {Gaussian} states},
  journal       = {Phys. Rev. A},
  volume        = {73},
  pages         = {033821},
  year          = {2006},
  doi           = {10.1103/PhysRevA.73.033821},
}

@article{Rivas2012,
  author        = {Rivas, {\'A}ngel and Luis, Alfredo},
  title         = {Sub-{Heisenberg} estimation of non-random phase shifts},
  journal       = {New J. Phys.},
  volume        = {14},
  pages         = {093052},
  year          = {2012},
  doi           = {10.1088/1367-2630/14/9/093052},
}

@misc{guo2025metrological,
  author        = {Guo, Jiajie and Liu, Shuheng and Jing, Boxuan and He, Qiongyi and Gessner, Manuel},
  title         = {Metrological sensitivity beyond {Gaussian} limits with cubic phase states},
  year          = {2025},
  eprint        = {2512.03769},
  archivePrefix = {arXiv},
  primaryClass  = {quant-ph},
}

@misc{Park2025,
  author        = {Park, Kimin and Krisnanda, Tanjung and Gao, Yvonne and Filip, Radim},
  title         = {Quantum phase estimation beyond the {Gaussian} limit},
  year          = {2025},
  eprint        = {2508.13046},
  archivePrefix = {arXiv},
  primaryClass  = {quant-ph},
}

@misc{gordillo2026quantum,
  author        = {Gordillo-Hachuel, Rub{\'e}n and Torrontegui, Erik and de Dios, Cristina and Puebla, Ricardo},
  title         = {Quantum metrological advantage of high-order squeezed states},
  year          = {2026},
  eprint        = {2604.09958},
  archivePrefix = {arXiv},
  primaryClass  = {quant-ph},
}

@article{Branford2021,
  author        = {Branford, Dominic and Rubio, Jes{\'u}s},
  title         = {Average number is an insufficient metric for interferometry},
  journal       = {New J. Phys.},
  volume        = {23},
  pages         = {123041},
  year          = {2021},
  doi           = {10.1088/1367-2630/ac3571},
}

@article{Braunstein94,
  author        = {Braunstein, Samuel L. and Caves, Carlton M.},
  title         = {Statistical distance and the geometry of quantum states},
  journal       = {Phys. Rev. Lett.},
  volume        = {72},
  pages         = {3439--3443},
  year          = {1994},
  doi           = {10.1103/PhysRevLett.72.3439},
}

@article{16Vahlbruch,
  author        = {Vahlbruch, Henning and Mehmet, Moritz and Danzmann, Karsten and Schnabel, Roman},
  title         = {Detection of 15 {dB} squeezed states of light and their application for the absolute calibration of photoelectric quantum efficiency},
  journal       = {Phys. Rev. Lett.},
  volume        = {117},
  pages         = {110801},
  year          = {2016},
  doi           = {10.1103/PhysRevLett.117.110801},
}

@article{24Chinni,
  author        = {Chinni, Karthik and Quesada, Nicol{\'a}s},
  title         = {Beyond the parametric approximation: Pump depletion, entanglement, and squeezing in macroscopic down-conversion},
  journal       = {Phys. Rev. A},
  volume        = {110},
  pages         = {013712},
  year          = {2024},
  doi           = {10.1103/PhysRevA.110.013712},
}

@article{70Walls,
  author        = {Walls, Daniel F. and Barakat, Richard},
  title         = {Quantum-mechanical amplification and frequency conversion with a trilinear {Hamiltonian}},
  journal       = {Phys. Rev. A},
  volume        = {1},
  pages         = {446--453},
  year          = {1970},
  doi           = {10.1103/PhysRevA.1.446},
}

@article{16Harder,
  author        = {Harder, Georg and Bartley, Tim J. and Lita, Adriana E. and Nam, Sae Woo and Gerrits, Thomas and Silberhorn, Christine},
  title         = {Single-mode parametric-down-conversion states with 50 photons as a source for mesoscopic quantum optics},
  journal       = {Phys. Rev. Lett.},
  volume        = {116},
  pages         = {143601},
  year          = {2016},
  doi           = {10.1103/PhysRevLett.116.143601},
}

@article{99Lloyd,
  author        = {Lloyd, Seth and Braunstein, Samuel L.},
  title         = {Quantum computation over continuous variables},
  journal       = {Phys. Rev. Lett.},
  volume        = {82},
  pages         = {1784--1787},
  year          = {1999},
  doi           = {10.1103/PhysRevLett.82.1784},
}

@article{06Wu,
  author        = {Wu, Re-Bing and Tarn, Tzyh-Jong and Li, Chun-Wen},
  title         = {Smooth controllability of infinite-dimensional quantum-mechanical systems},
  journal       = {Phys. Rev. A},
  volume        = {73},
  pages         = {012719},
  year          = {2006},
  doi           = {10.1103/PhysRevA.73.012719},
}

@article{Popoviciu1935,
  author        = {Popoviciu, Tiberiu},
  title         = {Sur les {\'e}quations alg{\'e}briques ayant toutes leurs racines r{\'e}elles},
  journal       = {Mathematica (Cluj)},
  volume        = {9},
  pages         = {129--145},
  year          = {1935},
}

@article{Pinel2013,
  author        = {Pinel, O. and Jian, P. and Treps, N. and Fabre, C. and Braun, D.},
  title         = {Quantum parameter estimation using general single-mode {Gaussian} states},
  journal       = {Phys. Rev. A},
  volume        = {88},
  pages         = {040102(R)},
  year          = {2013},
  doi           = {10.1103/PhysRevA.88.040102},
}

@article{Marian1993,
  author        = {Marian, Paulina and Marian, Tudor A.},
  title         = {Squeezed states with thermal noise. {I}. {P}hoton-number statistics},
  journal       = {Phys. Rev. A},
  volume        = {47},
  pages         = {4474--4486},
  year          = {1993},
  doi           = {10.1103/PhysRevA.47.4474},
}

@article{Fitzke2023,
  author        = {Fitzke, Erik and Niederschuh, Florian and Walther, Thomas},
  title         = {Simulating the photon statistics of multimode {Gaussian} states by automatic differentiation of generating functions},
  journal       = {APL Photonics},
  volume        = {8},
  pages         = {026106},
  year          = {2023},
  doi           = {10.1063/5.0129638},
}

@article{12Jarzyna,
  author        = {Jarzyna, Marcin and Demkowicz-Dobrza{\'n}ski, Rafa{\l}},
  title         = {Quantum interferometry with and without an external phase reference},
  journal       = {Phys. Rev. A},
  volume        = {85},
  pages         = {011801(R)},
  year          = {2012},
  doi           = {10.1103/PhysRevA.85.011801},
}

@article{Lee2019,
  author        = {Lee, Changhyoup and Oh, Changhun and Jeong, Hyunseok and Rockstuhl, Carsten and Lee, Su-Yong},
  title         = {Using states with a large photon number variance to increase quantum {Fisher} information in single-mode phase estimation},
  journal       = {J. Phys. Commun.},
  volume        = {3},
  pages         = {115008},
  year          = {2019},
  doi           = {10.1088/2399-6528/ab524a},
}

@article{96Law,
  title = {Arbitrary Control of a Quantum Electromagnetic Field},
  author = {Law, C. K. and Eberly, J. H.},
  journal = {Phys. Rev. Lett.},
  volume = {76},
  issue = {7},
  pages = {1055--1058},
  numpages = {0},
  year = {1996},
  month = {Feb},
  publisher = {American Physical Society},
  doi = {10.1103/PhysRevLett.76.1055},
  url = {https://link.aps.org/doi/10.1103/PhysRevLett.76.1055}
}

@article{09Hofheinz,
  author  = {Hofheinz, Max and Wang, H. and Ansmann, M. and Bialczak, Radoslaw C. and Lucero, Erik and Neeley, M. and O'Connell, A. D. and Sank, D. and Wenner, J. and Martinis, John M. and Cleland, A. N.},
  title   = {Synthesizing arbitrary quantum states in a superconducting resonator},
  journal = {Nature},
  volume  = {459},
  pages   = {546--549},
  year    = {2009},
  doi     = {10.1038/nature08005}
}

@article{Luis2017,
  title = {Breaking the weak Heisenberg limit},
  volume = {95},
  ISSN = {2469-9934},
  url = {http://dx.doi.org/10.1103/PhysRevA.95.032113},
  DOI = {10.1103/physreva.95.032113},
  number = {3},
  journal = {Physical Review A},
  publisher = {American Physical Society (APS)},
  author = {Luis,  Alfredo},
  year = {2017},
  month = Mar 
}
\newpage
\onecolumngrid
\appendix

\section{On the heavy-tailed states}
\label{Appx: Heavy-tailed states}
In this appendix, we show how to construct heavy-tailed probes with infinite QFI even at a finite mean photon number. To this end, we require a divergent photon-number variance while preserving a finite mean photon number.

First, we impose a finite mean photon number,
\begin{align}
    \ket{\psi_\gamma} = \mathcal{N}\sum_{n=1}^\infty\frac{1}{n^\gamma}\ket{n}   \implies  \bra{\psi_\gamma}\hat{n}\ket{\psi_\gamma} = \mathcal{N}^2\sum_{n=1}^\infty\frac{1}{n^{2\gamma-1}}.
\end{align}
The series $\sum_{r=1}^\infty r^{-\mu}$ converges only if $\mu>1$, which here requires $2\gamma-1>1$, that is, $\gamma > 1$.
Second, we require a divergent photon-number variance,
\begin{equation}
\bra{\psi_\gamma}\hat{n}^2\ket{\psi_\gamma} = \mathcal{N}^2\sum_{n=1}^\infty\frac{1}{n^{2\gamma-2}},
\end{equation}
which diverges for $2\gamma-2\leq 1$, that is, $\gamma\leq\frac{3}{2}$.
Hence, the expansion in Eq.~\eqref{Eq: Heavy-Tailed distributions} with exponent $\gamma\in (1,\frac{3}{2}]$ yields a finite mean photon number and a divergent QFI.

\section{On the details of asymmetrically superposed states}
\label{Appx: Resources}
In this appendix, we study how the weight parameter $\epsilon$ in Eq.~\eqref{Eq: Asymmetric Superposition} affects the QFI for a given excitation $\ket{\phi}$.

Since the QFI in Eq.~\eqref{Eq: QFI assymetric superposition} is concave in $\epsilon^2$, setting its derivative with respect to $\epsilon^2$ to zero gives the maximum at
\begin{equation}
    \epsilon_*^2=\min\left \{1,\frac{\langle\hat{n}^2\rangle_\phi}{2\langle\hat{n}\rangle_\phi^2}\right\}.
\end{equation}
If $\langle\hat{n}^2\rangle_\phi\geq 2\langle\hat{n}\rangle_\phi^2$ (excitations with broad number distributions), the optimal value is $\epsilon_*=1$ and the superposition reduces to $\ket{\phi}$. Whenever $\langle\hat{n}^2\rangle_\phi\approx\langle\hat{n}\rangle_\phi^2$, exact for $\ket{\phi}=\ket{N}$, the optimum lies at $\epsilon_*\approx2^{-1/2}$, the balanced superposition. Since $\langle\hat n^2\rangle_\phi\geq\langle\hat n\rangle_\phi^2$, we have $1/2\leq\epsilon_*^2\leq1$ for every excitation, so the optimum never lies in the $\epsilon\to0$ regime that appears when $\langle\hat n\rangle$ is taken as the resource.

\section{On the distribution attaining the RMS photon budget}\label{app_Existence_uniqueness_RMS}
In this appendix, we construct the distribution $t\in\mathcal Q_g$ attaining the maximum in the RMS photon budget $N_{\mathrm{RMS}}(g)$, defined in Eq.~\eqref{def_rms_resource}. We also show that the same distribution maximizes the mean photon number. For simplicity, we use the $N_{\max}=1$ notation, writing the envelope constraints as $q_n\leq g_n$ for $n\geq1$ and leaving the vacuum population unconstrained. An arbitrary free window is included by setting $g_n=1$ for $1\leq n\leq N_{\max}$, so the argument remains general. Throughout, we assume $\sum_{n\geq1}n^2g_n<\infty$.

First, consider the case $\sum_{n\geq 1}g_n\leq1$. Take the distribution $t$ as
\begin{equation}
 t_n=g_n\quad(n\geq1),\qquad
 t_0=1-\sum_{n\geq1}g_n.
 \label{eq:c-r-low}
\end{equation}
This distribution is normalized and admissible. For any
$q\in\mathcal Q_g$ and $k\in\{1,2\}$,
\begin{equation}
 \sum_{n\geq0}n^k t_n-\sum_{n\geq0}n^k q_n
 =\sum_{n\geq1}n^k(g_n-q_n)\geq0.
 \label{eq:c-low-proof}
\end{equation}
Thus, $t$ maximizes both the first and second moments, with $\bar N(g)=\sum_{n\geq1}ng_n$ and $N_{\mathrm{RMS}}^2(g)=\sum_{n\geq1}n^2g_n$. Equality requires $q_n=g_n$ for every $n\geq1$, while normalization fixes
$q_0=t_0$.

Now consider $\sum_{n\geq1}g_n>1$. Choose an integer $m\geq1$ satisfying
\begin{equation}\label{appx_t_sandwich}
 \sum_{n>m}g_n\leq1\leq\sum_{n\geq m}g_n.
\end{equation}
To show that such an $m$ exists, define the tail mass
$T(k)\coloneq \sum_{n\geq k}g_n$. It is non-increasing in $k$, it starts at
$T(1)=\sum_{n\geq 1}g_n>1$, and $T(k)\to 0$ as $k\to\infty$ since
$\sum_{n\geq 1}g_n<\infty$. Hence, the set
$\{k\geq 1 : T(k)\geq 1\}$ is non-empty and bounded, and its largest
element $m$ satisfies the previous inequality, i.e.,
$T(m)\geq1\geq T(m+1)$.
Next, take the distribution $t$ as
\begin{equation}\label{caseII_optimal_t_appx}
 t_n=\begin{cases}
  0,&0\leq n<m,\\[2pt]
  1-\displaystyle\sum_{j>m}g_j,&n=m,\\[4pt]
  g_n,&n>m.
 \end{cases}
\end{equation}
The distribution fills the envelope above $m$ and places the remaining probability at $m$. It is normalized and admissible, since Eq.~\eqref{appx_t_sandwich} ensures $0\leq t_m\leq g_m$. For any $q\in\mathcal Q_g$, normalization gives $\sum_{n\geq0}(t_n-q_n)=0$. Hence, for $k\in\{1,2\}$,
\begin{align}
 \sum_{n\geq0}n^k(t_n-q_n)
 &=\sum_{n\geq0}(n^k-m^k)(t_n-q_n)\notag\\
 &=\sum_{n>m}(n^k-m^k)(g_n-q_n)+\sum_{0\leq n<m}(m^k-n^k)q_n\geq0.
 \label{eq:c-high-proof}
\end{align}
Both sums are non-negative. Therefore, $t$ again maximizes both moments,
giving
\begin{equation}\label{appx_t_MPN}
\begin{split}
    \bar N(g)=\max_{q\in\mathcal Q_g}\mathbb E_q[\hat n]
  =\sum_{n\geq0}nt_n
  &=m+\sum_{n>m}(n-m)g_n,
\end{split}
\end{equation}
\begin{equation}\label{appx_t_RMSPN}
    N_{\mathrm{RMS}}^2(g)=\sum_{n\geq0}n^2t_n
  =m^2+\sum_{n>m}(n^2-m^2)g_n.
\end{equation}
Equality in Eq.~\eqref{eq:c-high-proof} requires $q_n=g_n$ for $n>m$ and $q_n=0$ for $n<m$; normalization then fixes $q_m=t_m$. Thus, in either case, only $q=t$ attains equality.

\section{On the algebraically decaying envelope example}\label{Appx:algebraic_ex}
In this appendix, we detail the computations for the algebraically decaying envelope example in Sec.~\ref{Secc_NG_advg}. We derive the asymptotic behavior of $\bar N(g)$ and $N_{\mathrm{RMS}}(g)$ for the envelope of Eq.~\eqref{algebraic_env_ex} and obtain the limit in Eq.~\eqref{alg_ex_limit_NG_adv}. Throughout, $p>3$ is fixed and $o(1)$ denotes terms that vanish as $N_{\max}\to\infty$. Note that for $n>N_{\max}$ we have $n^2g_n=N_{\max}^p\,n^{2-p}$, so the condition $\sum_{n\geq N_\mathrm{max}}n^2g_n<\infty$ holds precisely because $p>3$.

We first identify the distribution $t$ attaining both $\bar N(g)$ and $N_{\mathrm{RMS}}(g)$. Since $g_1=1$ and the tail of the envelope is strictly positive, $\sum_{n\ge1}g_n>1$, so $t$ is given by Eq.~\eqref{caseII_optimal_t_appx}, with $m$ the largest integer satisfying $T(m)\ge1$. The last level of the free window carries $g_{N_{\max}}=1$; hence,
\begin{equation}
  T(N_{\max})=1+\sum_{n>N_{\max}}g_n>1,
\end{equation}
and therefore $m\ge N_{\max}$. Thus, every level $n>m$ belongs to the
algebraic tail, where $g_n=N_{\max}^p\,n^{-p}$, and also
$g_m=(N_{\max}/m)^p$, including at $m=N_{\max}$, where both sides equal one.

From Eqs.~\eqref{appx_t_MPN} and \eqref{appx_t_RMSPN}, we can write $\bar N(g)$ and $N_{\mathrm{RMS}}(g)$ in terms of $m$ and the tail sums
\begin{equation}\label{eq:D_sigma}
  S_j:=\sum_{n>m}n^j g_n=N_{\max}^p\sum_{n>m}n^{\,j-p},\quad j=0,1,2.
\end{equation}
Indeed, splitting $\sum_{n>m}(n^k-m^k)g_n=S_k-m^kS_0$, both moments read
\begin{equation}\label{eq:D_budgets}
\begin{aligned}
  \bar N(g)&=S_1+m\,(1-S_0),\\
  N^2_{\mathrm{RMS}}(g)&=S_2+m^2(1-S_0).
\end{aligned}
\end{equation}
To estimate the tail sums, note that for $j\le2<p-1$ the function $f(x)=x^{j-p}$ is decreasing, so that $\int_n^{n+1}f\le f(n)\le\int_{n-1}^{n}f$. Summing over $n\ge m+1$ and using $\int_k^\infty x^{j-p}dx=k^{\,j+1-p}/(p-j-1)$, we obtain
\begin{equation}\label{eq:D_sandwich}
  \frac{N_{\max}^p\,(m+1)^{j+1-p}}{p-j-1}\leq S_j\le
  \frac{N_{\max}^p\,m^{j+1-p}}{p-j-1}.
\end{equation}
The two bounds differ by the factor $\bigl(m/(m+1)\bigr)^{p-j-1}=1-(p-j-1)/m+\mathcal{O}(m^{-2})$, which approaches one since $m\ge N_{\max}\to\infty$. Hence,
\begin{equation}\label{eq:D_sigma_asym}
  S_j=\frac{N_{\max}^p\,m^{j+1-p}}{p-j-1}\,\left(1+\mathcal{O}\left(\frac{1}{m}\right)\right)
\end{equation}

Next, we determine how $m$ grows with $N_\mathrm{max}$. Since $T(m+1)=S_0$ and $T(m)=S_0+g_m$, Eq.~\eqref{appx_t_sandwich} reads
\begin{equation}\label{eq:D_pivot}
  1-g_m\leq S_0\le1.
\end{equation}
Combining $S_0\le1$ with the lower bound in Eq.~\eqref{eq:D_sandwich} for $j=0$ gives $(m+1)^{p-1}\ge N_{\max}^p/(p-1)$, that is,
\begin{equation}
     \frac{m+1}{N_{\max}}\ge\Bigl(\frac{N_{\max}}{p-1}\Bigr)^{1/(p-1)}
     \xrightarrow{\,N_{\max}\to\infty\,}\infty .
\end{equation}

Thus, $m/N_{\max}\to\infty$ and $g_m=(N_{\max}/m)^p\to0$. Then,
Eq.~\eqref{eq:D_pivot} yields $S_0\to1$. Moreover, the integral bounds on $T(m)$ and $T(m+1)$ give
$\left|m-(N_{\max}^p/(p-1))^{1/(p-1)}\right|\leq1$,
which sharpens the asymptotic corrections. Eq.~\eqref{eq:D_sigma_asym} with $j=0$ then fixes $m$,
\begin{equation}\label{eq:D_m}
\begin{aligned}
  N_{\max}^p&=(p-1)m^{p-1}\bigl(1+O(m^{-1})\bigr),\\
  m&=(p-1)^{-\frac{1}{p-1}}\,N_{\max}^{\frac{p}{p-1}}
  \bigl(1+O(N_{\max}^{-\frac{p}{p-1}})\bigr).
\end{aligned}
\end{equation}
Writing $N_{\max}^p m^{j+1-p}=m^j\,N_{\max}^p m^{1-p}$ and inserting the first line of Eq.~\eqref{eq:D_m} into Eq.~\eqref{eq:D_sigma_asym}, we find
\begin{equation}\label{eq:D_sigma12}
\begin{aligned}
S_1&=\frac{p-1}{p-2}\,m\bigl(1+O(m^{-1})\bigr),\\
  S_2&=\frac{p-1}{p-3}\,m^2\bigl(1+O(m^{-1})\bigr).
\end{aligned}
\end{equation}

We can now evaluate $\bar N(g)$ and $N_{\mathrm{RMS}}(g)$. By Eq.~\eqref{eq:D_pivot}, $0\le1-S_0\le g_m=O(m^{-1})$, so the terms $m(1-S_0)$ and $m^2(1-S_0)$ in Eq.~\eqref{eq:D_budgets} give relative corrections of order $O(m^{-1})$ compared with $S_1\propto m$ and $S_2\propto m^2$. Therefore,
\begin{equation}\label{eq:D_budgets_asym}
\begin{aligned}
  \bar N(g)&=\frac{p-1}{p-2}\,m\bigl(1+O(m^{-1})\bigr),\\
  N^2_{\mathrm{RMS}}(g)&=\frac{p-1}{p-3}\,m^2\bigl(1+O(m^{-1})\bigr),
\end{aligned}
\end{equation}
and, using Eq.~\eqref{eq:D_m}, we see that both quantities scale as
$N_{\max}^{p/(p-1)}$ for $N_\mathrm{max}\gg1$,
\begin{equation}
\begin{aligned}
  \bar N(g)&=\frac{(p-1)^{\frac{p-2}{p-1}}}{p-2}\,N_{\max}^{\frac{p}{p-1}}
  \bigl(1+O(N_{\max}^{-\frac{p}{p-1}})\bigr),\\
  N_{\mathrm{RMS}}(g)&=\frac{(p-1)^{\frac{p-3}{2(p-1)}}}{\sqrt{p-3}}\,N_{\max}^{\frac{p}{p-1}}
  \bigl(1+O(N_{\max}^{-\frac{p}{p-1}})\bigr).
\end{aligned}
\end{equation}
This growth, faster than $N_{\max}$, reflects that $t$ leaves the whole free window empty and places its weight deep in the tail, since $m/N_{\max}\to\infty$. Since $t_n=0$ for $n<m$, we have $\bar N(g)\ge m\ge N_{\max}$, so $1/\bar N(g)\to0$ and $\bar N(g)+1=\bar N(g)\,[1+O(m^{-1})]$.
 
Inserting Eq.~\eqref{eq:D_budgets_asym} into the right-hand side of Eq.~\eqref{bound_NG_resource}, the factors $m^2$ cancel and
\begin{equation}\label{eq:D_limit}
\begin{aligned}
  \frac{N^2_{\mathrm{RMS}}(g)}{8\bar N(g)\,[\bar N(g)+1]}
  &=\frac{(p-1)/(p-3)}{8\,(p-1)^2/(p-2)^2}\,
  \bigl[1+O(N_{\max}^{-\frac{p}{p-1}})\bigr]\\
  &\simeq\frac{(p-2)^2}{8(p-1)(p-3)},
\end{aligned}
\end{equation}
which is Eq.~\eqref{alg_ex_limit_NG_adv} for $N_\mathrm{max}\gg 1$. This limit exceeds one when $(p-2)^2>8(p-1)(p-3)$. Writing $u=p-2>1$, so that $(p-1)(p-3)=u^2-1$, the condition becomes $u^2>8(u^2-1)$, i.e., $u^2<8/7$. Hence, Eq.~\eqref{bound_NG_resource} certifies a non-Gaussian advantage for $3<p<2+\sqrt{8/7}\approx3.069$ and sufficiently large $N_{\max}$.

\section{On the bound on the QFI for Gaussian states in terms of tail decay}\label{Appx_sec_bound_tail_decay}
In this appendix, we show that every admissible pure Gaussian probe satisfies $\tanh r\le\rho_g$, and we derive the Gaussian bound in Eq.~\eqref{bound_gaussian_qfi_tail_decay}. Throughout, $0<\rho_g<1$.
 
Up to a global phase, every pure Gaussian state can be written as $|\psi_G\rangle=D(\alpha)S(\xi)|0\rangle$ with $\xi=re^{i\varphi}$ and $r\ge0$~\cite{Monras2006,Pinel2013}. A phase rotation $e^{-i\phi\hat n}$ leaves the photon-number distribution unchanged and only shifts the phases of $\alpha$ and $\xi$, so we may take $\xi=r$ real. We write $n_r:=\sinh^2r$, $\beta:=\tanh r\in[0,1)$, and $\alpha=u+iv$, and split the displacement along the anti-squeezed and squeezed quadratures, $|\alpha|^2=d_++d_-$ with $d_+=v^2$ and $d_-=u^2$ (for real $\xi$, $\hat S(\xi)$ squeezes the real quadrature, as in Eq.~\eqref{eq:G_sqvac}). The photon-number mean
and variance are then~\cite{Monras2006,Pinel2013}
\begin{equation}\label{app_moments_tail_decay}
\begin{aligned}
  \langle\hat n\rangle_{\psi_G}&=n_r+d_++d_-,\\
  \mathrm{Var}_{\psi_G}(\hat n)&=2n_r(n_r+1)+e^{2r}d_++e^{-2r}d_- .
\end{aligned}
\end{equation}
To compare the photon-number distribution
$q_n=|\langle n|\psi_G\rangle|^2$ with the envelope, we use its
photon-number generating function,
\begin{equation}\label{eq:Sq_Gdef}
  G(z):=\sum_{n\ge0}q_n\,z^n=\langle\psi_G|\,z^{\hat n}\,|\psi_G\rangle,
  \qquad z\ge0 .
\end{equation}
This is a power series whose coefficients are the photon-number probabilities. It satisfies $G(1)=1$, and its derivatives at $z=1$ give the factorial moments of $\hat n$. Here, we only need its radius of convergence $R$, which by the Cauchy--Hadamard formula satisfies $1/R=\limsup_{n\to\infty}q_n^{1/n}$. The radius of convergence therefore measures how fast $q_n$ decays at large photon numbers, in the same way as $\rho_g$ in Eq.~\eqref{defs_tail_decay} measures the decay of the envelope. Comparing $R$ with $1/\rho_g$ thus compares the tail of a Gaussian distribution with the tail allowed by the envelope. For a pure Gaussian state, $G$ is known in closed form~\cite{Marian1993,Fitzke2023},
\begin{equation}\label{eq:Sq_G}
\begin{aligned}
  G(z)&=\sqrt{\frac{1-\beta^2}{1-\beta^2z^2}}\exp\Bigl[\frac{d_+(1-\beta)(z-1)}{1-\beta z}
  +\frac{d_-(1+\beta)(z-1)}{1+\beta z}\Bigr],
\end{aligned}
\end{equation}
where the series converges for $0\le z<1/\beta$.
 
We first determine the radius of convergence of $G$. Suppose first that $\beta>0$. For $1\le z<1/\beta$, both terms in the exponent of Eq.~\eqref{eq:Sq_G} are non-negative; hence,
\begin{equation}
  G(z)\ge\sqrt{\frac{1-\beta^2}{1-\beta^2z^2}} .
\end{equation}
The right-hand side diverges as $z$ approaches $1/\beta$ from below. Thus, the series converges below $1/\beta$ but not at $1/\beta$, so its radius of convergence is exactly $R=1/\beta$, and the Cauchy--Hadamard formula gives
\begin{equation}\label{eq:Sq_rate}
  \limsup_{n\to\infty}q_n^{1/n}=\beta=\tanh r ,
\end{equation}
i.e., the photon-number distribution of a pure Gaussian state decays at the exponential rate $\ln\coth r$, in agreement with $\kappa$ in Eq.~\eqref{tail_decay_implementation}. An admissible probe satisfies $q_n\le g_n$ for $n>N_{\max}$; hence, $q_n^{1/n}\le g_n^{1/n}$. Taking the limit superior and using Eq.~\eqref{defs_tail_decay},
\begin{equation}\label{appx_limit_squeezing_tail}
  \tanh r\le\rho_g ,
\end{equation}
which also holds trivially for $\beta=0$. This restriction holds independently of the displacement and of $N_{\max}$, since it only involves the asymptotic decay of the envelope.
 
Next, we derive an upper bound on the QFI of Gaussian states in terms of the tail decay. The functions $x^2/(1-x^2)$ and $(1+x)/(1-x)$ are increasing on $0\le x<1$, and $n_r=\beta^2/(1-\beta^2)$, $e^{2r}=(1+\beta)/(1-\beta)$.
Hence,
\begin{equation}\label{eq:Sq_sg}
  n_r\le n_g:=\frac{\rho_g^2}{1-\rho_g^2},\qquad
  e^{2r}\le\tau_g:=\frac{1+\rho_g}{1-\rho_g}.
\end{equation}
Moreover, since the photon-number distribution of $\ket{\psi_G}$ is admissible, Eq.~\eqref{app_moments_tail_decay} gives $d_++d_-\le\langle\hat n\rangle_{\psi_G}\le\bar N(g)$. Inserting these bounds into the QFI for phase estimation, Eq.~\eqref{Eq: QFI from generator n}, and using $e^{-2r}\le e^{2r}$, we obtain
\begin{equation}\label{appx_qfi_tail_decay}
\begin{aligned}
  \mathcal{F}_\theta(\psi_G)&\le8n_r(n_r+1)+4e^{2r}(d_++d_-)\\
  &\le8n_g(n_g+1)+4\tau_g\,\bar N(g) .
\end{aligned}
\end{equation}
The first term is the squeezing contribution and the second is the displacement contribution.
 
Finally, we express these constants through the tail length. Writing $\rho_g=e^{-1/\ell_g}$ and using the elementary inequalities $\sinh x\ge x$ and $e^x-1\ge x$,
\begin{equation}\label{appx_bounds_tail}
\begin{aligned}
  8n_g(n_g+1)&=\frac{8\rho_g^2}{(1-\rho_g^2)^2}
  =\frac{2}{\sinh^2(1/\ell_g)}\le2\ell_g^2,\\
  \tau_g&=1+\frac{2}{e^{1/\ell_g}-1}\le1+2\ell_g .
\end{aligned}
\end{equation}
Inserting the bounds in Eq.~\eqref{appx_bounds_tail} into Eq.~\eqref{appx_qfi_tail_decay} and using $\bar N(g)\le N_{\mathrm{RMS}}(g)$, which holds because $t$ attains both budgets and $\mathbb{E}_t[\hat{n}]^2\leq\mathbb{E}_t[\hat{n}^2]=N_{\mathrm{RMS}}^2(g)$, we obtain
\begin{equation}\label{eq:Sq_final}
\begin{aligned}
  \mathcal{F}_\theta(\psi_G)&\le2\ell_g^2+4\,(1+2\ell_g)\,\bar N(g)\\
  &\le2\ell_g^2+4\,(1+2\ell_g)\,N_{\mathrm{RMS}}(g),
\end{aligned}
\end{equation}
which is Eq.~\eqref{bound_gaussian_qfi_tail_decay}.

\section{On the exponentially decaying envelope example}\label{Appx:exponential_ex}
In this appendix, we detail the computations for the exponentially decaying envelope example in Sec.~\ref{Secc_NG_advg}, Eq.~\eqref{exp_env_ex}. In particular, we compute both $N_\mathrm{RMS}(g)$ and $\bar{N}(g)$, evaluate both criteria, Eqs.~\eqref{bound_NG_resource} and \eqref{bound_NG_decay}, and show that the best admissible Gaussian probe has a QFI linear in $N_{\mathrm{RMS}}(g)$.

We first identify the distribution $t$ attaining both $\bar N(g)$ and $N_{\mathrm{RMS}}(g)$. Since $g_1=1$ and the tail of the envelope is strictly positive, $\sum_{n\ge1}g_n>1$, so $t$ is given by Eq.~\eqref{caseII_optimal_t_appx}, with $m$ the largest integer satisfying $T(m)\ge1$. The tail masses just above the free window are
\begin{equation}
\begin{split}
    T(N_{\max}+1)&=\sum_{j\ge1}2^{-j}=1,\\
    T(N_{\max}+2)&=\frac12;
\end{split}
\end{equation}
hence, $m=N_{\max}+1$. The remaining probability placed at $m$ is $t_m=1-\sum_{n>m}g_n=1/2=g_m$, so $t$ fills the envelope exactly above $N_{\max}$,
\begin{equation}\label{eq:E_t}
  \begin{split}
      t_{N_{\max}+j}&=2^{-j}\ \ (j\ge1),\\ 
  t_n&=0\ \ (0\le n\le N_{\max}).
  \end{split}
\end{equation}
The free window, including the vacuum, is left unoccupied, and all the probability lies in the tail.

To evaluate the moments of $t$, we use the geometric sums $\sum_{j\ge1}x^j=x/(1-x)$, $\sum_{j\ge1}j\,x^j=x/(1-x)^2$, and $\sum_{j\ge1}j^2x^j=x(1+x)/(1-x)^3$, obtained by differentiating the first one. At $x=1/2$, they give
\begin{equation}\label{eq:E_sums}
  \sum_{j\ge1}2^{-j}=1,\qquad \sum_{j\ge1}j\,2^{-j}=2,\qquad
  \sum_{j\ge1}j^2\,2^{-j}=6 .
\end{equation}
Expanding $(N_{\max}+j)^k$ for $k=1,2$ and using Eq.~\eqref{eq:E_sums},
\begin{equation}\label{eq:E_budgets}
\begin{split}
  \bar N(g)&=\sum_{j\ge1}(N_{\max}+j)\,2^{-j}=N_{\max}+2,\\
  N^2_{\mathrm{RMS}}(g)&=\sum_{j\ge1}(N_{\max}+j)^2\,2^{-j}=N_{\max}^2+4N_{\max}+6.
\end{split}
\end{equation}
Since $(N_{\max}+2)^2\le N^2_{\mathrm{RMS}}(g)\le (N_{\max}+3)^2$, the RMS budget and the threshold differ only by a constant,
\begin{equation}\label{eq:E_window}
  N_{\max}+2\le N_{\mathrm{RMS}}(g)\le N_{\max}+3 .
\end{equation}
Moreover, Eq.~\eqref{QFI_lower_bound_best_NG} gives the QFI of the probe $|\psi_t\rangle$ exactly,
\begin{equation}\label{eq:E_Ft}
  \mathcal{F}_\theta(\psi_t)=2N^2_{\mathrm{RMS}}(g)-\bar N^2(g)
  =N_{\max}^2+4N_{\max}+8 .
\end{equation}

Next, we evaluate the mean-based criterion. Inserting Eq.~\eqref{eq:E_budgets} into the right-hand side of Eq.~\eqref{bound_NG_resource}, we obtain
\begin{equation}\label{eq:E_mean}
  \frac{N^2_{\mathrm{RMS}}(g)}{8\bar N(g)\,[\bar N(g)+1]}
  =\frac{N_{\max}^2+4N_{\max}+6}{8\,(N_{\max}^2+5N_{\max}+6)}
  <\frac18 ,
\end{equation}
where the inequality holds for every $N_{\max}\ge1$, and the left-hand side tends to $1/8$ as $N_{\max}\to\infty$. Thus, Eq.~\eqref{bound_NG_resource} never certifies an advantage for this envelope: the mean of $t$ grows as fast as its RMS value, so the mean-based Gaussian bound, Eq.~\eqref{QFI_gaussian_envelope_framework}, also grows quadratically.

We now turn to the tail-decay-based criterion, Eq.~\eqref{bound_NG_decay}. For $n>N_{\max}$, $g_n^{1/n}=2^{-(n-N_{\max})/n}=2^{-1+N_{\max}/n}$, which converges to $1/2$ as $n\to\infty$. Hence, the limit superior in Eq.~\eqref{defs_tail_decay} is a limit, and
\begin{equation}\label{eq:E_rho}
  \rho_g=\frac12,\qquad \ell_g=\frac{1}{\ln 2},
\end{equation}
independently of $N_{\max}$. Inserting these values into Eqs.~\eqref{bound_gaussian_qfi_tail_decay} and \eqref{bound_NG_decay}, we obtain
\begin{align}
  \mathcal{F}_\theta(\psi_G)&\le\frac{2}{\ln^22}
  +4\Bigl(1+\frac{2}{\ln2}\Bigr)N_{\mathrm{RMS}}(g),\label{eq:E_FG}\\
  \frac{\mathcal{F}_\theta(\psi_t)}{\mathcal{F}_\theta(\psi_G)}&\ge
  \frac{N^2_{\mathrm{RMS}}(g)}{2/\ln^22+4\,(1+2/\ln2)\,N_{\mathrm{RMS}}(g)} .
  \label{eq:E_ratio}
\end{align}
The right-hand side of Eq.~\eqref{eq:E_ratio} grows linearly with $N_{\mathrm{RMS}}(g)$. Numerically, the denominator is $4.16+15.54\,N_{\mathrm{RMS}}(g)$, so the ratio exceeds one for $N_{\mathrm{RMS}}(g)>15.805$, which Eq.~\eqref{eq:E_window} guarantees for $N_{\max}\ge14$.

Finally, we show that the Gaussian bound in Eq.~\eqref{eq:E_FG} is attained up to constants, so that the separation is genuinely one of scaling. Consider a coherent state $|\alpha\rangle$ with $A\coloneq|\alpha|^2\le N_{\max}/2$, whose photon-number distribution is $q_n=e^{-A}A^n/n!$. The levels $1\le n\le N_{\max}$ lie in the free window and impose no condition. For $n\ge1$, maximizing $e^{-A}A^n$ over $A$ (the maximum lies at $A=n$) and using Stirling's lower bound, $n!\ge\sqrt{2\pi n}\,(n/e)^n$, we obtain
\begin{equation}\label{eq:E_peak}
  q_n\le\frac{e^{-n}n^n}{n!}\le\frac{1}{\sqrt{2\pi n}}<\frac12 .
\end{equation}
For the tail levels, the ratio of consecutive populations is $q_{N_{\max}+k}/q_{N_{\max}+k-1}=A/(N_{\max}+k)\le1/2$ for $k\ge2$, since $A\le N_{\max}/2$. Combining this with Eq.~\eqref{eq:E_peak} at $n=N_{\max}+1$, we find, for every $j\ge1$,
\begin{equation}
  q_{N_{\max}+j}\le q_{N_{\max}+1}\,2^{-(j-1)}<2^{-j}=g_{N_{\max}+j},
\end{equation}
so the state is admissible. Its QFI is $\mathcal{F}_\theta=4\,\mathrm{Var}(\hat n)=4A$, and taking $A=N_{\max}/2$ gives $\mathcal{F}_\theta=2N_{\max}\ge2N_{\mathrm{RMS}}(g)-6$ by Eq.~\eqref{eq:E_window}. Together with Eq.~\eqref{eq:E_FG}, the best
admissible Gaussian probe, $\ket{\psi^*_\mathrm{G}}$, satisfies
\begin{equation}\label{eq:E_Gauss}
\begin{aligned}
  \mathcal{F}_\theta(\psi^*_G)&\ge2N_{\mathrm{RMS}}(g)-6,\\
  \mathcal{F}_\theta(\psi^*_G)&\le\frac{2}{\ln^22}
  +4\Bigl(1+\frac{2}{\ln2}\Bigr)N_{\mathrm{RMS}}(g),
\end{aligned}
\end{equation}
that is, $\mathcal{F}_\theta(\psi^*_G)\sim N_{\mathrm{RMS}}(g)$, whereas Eqs.~\eqref{eq:E_window} and \eqref{eq:E_Ft} give $\mathcal{F}_\theta(\psi_t)\sim N^2_{\mathrm{RMS}}(g)$. Enlarging the free window increases the RMS budget, but the fixed tail decay keeps $\tanh r\le1/2$, so admissible Gaussian probes can only profit from it through displacement, which yields a QFI linear in the resource.

\section{On the details of the example}
\label{Appx: Details Example}
In this appendix, we derive the results of the example in Sec.~\ref{Sec: Example}.
First, we compute the tail decay rate imposed by the setup.
Second, we find the threshold $N_{\rm max}$ that bounds the effective space in which the accessible states are defined.
Next, we establish the relation between $N_{\rm max}$ and $N_{\rm RMS}$ for this setup.
Finally, we derive the scaling of the quantum Fisher information with $N_{\rm max}$ for different classes of states.

\subsection{Shape of the envelope}
\label{Appx: Shape envelope}
To find the decay of the photon-number distribution generated by the universal gate set $\mathcal{G}$, we first compute the moment-generating function of the state,
\begin{align}
    M(s) = \sum_m q_m e^{sm}.
\end{align}
For a decay $q_m \propto e^{-\kappa m}$, $M(s)$ converges only for $s< \kappa$. 
Since every term of the sum is positive, each term is bounded by the whole sum, $q_me^{sm}\le M(s)$. Defining $N_s=s^{-1}\ln M(s)$, we can rewrite this bound as
\begin{align}
    q_{m} \le e^{-s(m-N_s)},
\end{align}
which holds for every $m$ and every $s$ at which $M(s)$ is finite. The supremum $\kappa$ of such $s$ gives the asymptotic decay rate of $q_m$. For a squeezed state, the coefficients are
\begin{align}
    c_{2n}=\frac{1}{\sqrt{\cosh r}}\,\frac{\sqrt{(2n)!}}{2^n n!}\,\big(-e^{i\vartheta}\tanh r\big)^n,\qquad c_{2n+1}=0 .\label{eq:G_sqvac}
\end{align}
The moment-generating function then reads
\begin{align}
    M(s)  &  =\frac{1}{\cosh r}\sum_n\binom{2n}{n}\frac{1}{4^n}\big(e^{2s}\tanh^2r\big)^n.
\end{align}
This series converges only if $s <\ln\coth r$, which fixes the supremum of the decay rate. The displacement multiplies the coefficients by a Hermite-polynomial factor~\cite{Marian1993} that grows more slowly than any exponential. This factor leaves the radius of convergence of $M(s)$, and hence $\kappa$, unchanged [Eq.~\eqref{eq:Sq_rate}]. When the displacement has a component $d_+>0$ along the anti-squeezed quadrature, however, the generating function in Eq.~\eqref{eq:Sq_G} diverges as $\exp[d_+(1-\beta)^2/\beta(1-\beta z)]$ for $z\to1/\beta$. A bound $q_n\leq C e^{-\kappa n}$ for all large $n$ would instead allow at most a simple pole, $G(z)\leq\sum_{n\leq N}q_nz^n+C(\beta z)^{N+1}/(1-\beta z)$. Consequently, no such bound with finite $C$ holds, and the decay rate of the envelope must satisfy $0<s<\kappa$ strictly. For any such $s$, $M(s)$ is finite and, for a displaced squeezed state, follows in closed form from Eq.~\eqref{eq:Sq_G} at $z=e^s$.

We therefore fix $0<s<\kappa$ and define $g_n=A\,e^{-s(n-N_\mathrm{max}-1)}$ for $n>N_\mathrm{max}$. Requiring that the probability weight above $N_{\rm max}$ equal a tolerance $\varepsilon$ gives
\begin{align}
    \sum_{n>N_\mathrm{max}}g_n=\frac{A}{1-e^{-s}}=\varepsilon
    \;\Rightarrow\; A = \varepsilon(1-e^{-s}).
\end{align}
Every probe then satisfies $q_n\leq M(s)e^{-sn}\leq g_n$ for $n>N_\mathrm{max}$, provided that $M(s)\,e^{-s(N_\mathrm{max}+1)}\leq\varepsilon(1-e^{-s})$.

\subsection{Threshold $N_{\rm max}$}
For Gaussian circuits, the threshold depends only on the displacement and squeezing gates. Each squeezing gate stretches phase-space distances by at most $e^{|r_i|}$, and rotations preserve them. The final squeezing therefore satisfies $|\xi'|\leq r_{\rm tot}$, and each displacement is amplified by at most $e^{r_{\rm tot}}$, so that the final displacement satisfies $|\alpha'|\leq e^{r_{\rm tot}}|\alpha_{\rm tot}|$. Since Eq.~\eqref{eq:Sq_G} increases with the squeezing and with the anti-squeezed displacement, the worst case is the aligned state.

Rotations and Kerr gates leave the photon-number distribution unchanged when they act, but they modify the phases on which subsequent squeezing and displacement operate, and hence the final distribution. For circuits with Kerr gates, we assume that this interleaving does not enlarge the envelope beyond that of the aligned Gaussian state with the same totals. A phase-space argument supports this assumption: Kerr gates preserve the angle-averaged Wigner function, squeezing enlarges phase-space distances by at most $e^{r}$, and displacements shift them by at most $|\alpha|$. This argument is heuristic, because Kerr gates generate Wigner negativity, and tail probabilities of quasiprobability distributions are not monotone under phase-space maps.
For circuits without squeezing, the assumption holds. Number-diagonal gates preserve the factorial moments $\|\hat a^k\psi\|^2=\langle\hat n(\hat n-1)\cdots(\hat n-k+1)\rangle$, and $\hat a\hat D(\gamma)=\hat D(\gamma)(\hat a+\gamma)$ gives $\|\hat a^k\hat D(\gamma)\phi\|\leq\sum_j\binom{k}{j}|\gamma|^{k-j}\|\hat a^j\phi\|$. By induction from the vacuum, $\|\hat a^k\psi\|\leq|\alpha_{\rm tot}|^k$, and expanding $G(z)$ in factorial moments gives, for $z\geq1$,
\begin{align}
    G(z)\leq e^{|\alpha_{\rm tot}|^2(z-1)},
\end{align}
the generating function of the coherent state $\ket{\alpha_{\rm tot}}$. Since this coherent state belongs to the Gaussian class, these circuits satisfy the same envelope.

For the purpose of computing $N_{\rm max}$, we can then describe the state-generation protocol as $D(\alpha_{\rm eff})S(-r_{\rm tot})\ket{0}$, with
\begin{align}
    \alpha_{\rm eff} \equiv e^{r_{\rm tot}}\alpha_{\rm tot} = e^{\sum_i |r_i|}\sum_j |\alpha_j|.
\end{align}
In the anti-squeezed quadrature $\hat x=(\hat a+\hat a^\dagger)/\sqrt2$, this state has mean $\mu=\sqrt2\,e^{r_{\rm tot}}\alpha_{\rm tot}$ and variance $\sigma^2=e^{2r_{\rm tot}}/2$, whereas the orthogonal quadrature has variance $e^{-2r_{\rm tot}}/2$. For strong anti-squeezing, $e^{2r_{\rm tot}}\gg1$, this quadrature dominates the photon number, $\hat n\approx\hat x^2/2$, and the tail probability is approximately
\begin{align}
    P(n>N_{\rm max})\approx\tfrac12\,\mathrm{erfc}\!\left(\frac{\sqrt{2N_{\rm max}}-\mu}{\sqrt2\,\sigma}\right)+\tfrac12\,\mathrm{erfc}\!\left(\frac{\sqrt{2N_{\rm max}}+\mu}{\sqrt2\,\sigma}\right).
\end{align}
The second term never exceeds the first. Using $\tfrac12\mathrm{erfc}(y)\simeq e^{-y^2}/(2\sqrt\pi\,y)$ for $y\gg1$ and setting the tail probability equal to $\varepsilon$, the argument of the first term satisfies $y^2=\ln\varepsilon^{-1}+\mathcal{O}(\ln\ln\varepsilon^{-1})$. To leading order in $\ln\varepsilon^{-1}$, this gives
\begin{align}
    N_\mathrm{max}\simeq e^{2r_{\rm tot}}\Big(|\alpha_{\rm tot}|+\sqrt{\tfrac12\ln\varepsilon^{-1}}\Big)^2 .
    \label{Eq: Appx - N_max example}
\end{align}
This quantile bounds only the total tail weight, not each level, and therefore does not guarantee Eq.~\eqref{tail_decay_implementation}. The certified threshold, Eq.~\eqref{Nmax_certified}, imposes the condition of Sec.~\ref{Appx: Shape envelope} with $M(s)$ given by Eq.~\eqref{eq:Sq_G} for the aligned state at $z=e^s$,
\begin{align}
    \ln M(s)=\frac12\ln\frac{1-\beta^2}{1-\beta^2e^{2s}}+e^{2r_{\rm tot}}|\alpha_{\rm tot}|^2\,\frac{(1-\beta)(e^s-1)}{1-\beta e^s}.
\end{align}
Since $N_{\rm RMS}$ diverges both as $s\to0$ and as $s\to\kappa$, it attains a minimum at some $s_*\in(0,\kappa)$. For large displacement, the resulting threshold remains close to Eq.~\eqref{Eq: Appx - N_max example}.

\subsection{Relation between $N_{\rm RMS}$ and $N_{\rm max}$}
To find $N_{\rm RMS}$, we maximize $\braket{n^2}$ under the envelope constraint using Appendix~\ref{app_Existence_uniqueness_RMS}. The tail masses are $T(N_{\rm max})=1+\varepsilon\geq1$ and $T(N_{\rm max}+1)=\varepsilon\leq1$, so the pivot is $m=N_{\rm max}$. The maximizing distribution therefore fills the tail completely, with total weight $\varepsilon$, and places the remaining probability $1-\varepsilon$ at $N_{\rm max}$. Writing $n=N_{\rm max}+J$ in the tail, with $x=e^{-s}$, the excess $J\geq1$ follows the normalized geometric law $h_J=(1-x)\,x^{J-1}$, whose first two moments are
\begin{align}
    \mathbb{E}[J]=\frac{1}{1-x},\qquad \mathbb{E}[J^2]=\frac{1+x}{(1-x)^2}.
\end{align}
The RMS photon budget then reads
\begin{align}
    N_\mathrm{RMS}^2 &= (1-\varepsilon)\,N_{\rm max}^2+\varepsilon\,\mathbb{E}\big[(N_{\rm max}+J)^2\big]
    \nonumber \\
    &= N_{\rm max}^2+\varepsilon\big(2N_{\rm max}\,\mathbb{E}[J]+\mathbb{E}[J^2]\big)
    \nonumber \\
    & =N_\mathrm{max}^2+\varepsilon\left[\frac{2N_\mathrm{max}}{1-e^{-s}}+\frac{1+e^{-s}}{(1-e^{-s})^2}\right].
\end{align}

\subsection{Scaling of the representative probes}
First, we compute the scaling of the ON state. This state has no probability weight above $N_{\rm max}$, so its variance and QFI scale with $N_{\rm max}$ as
\begin{align}
    \max_{N \in \{0,\dots,N_{\max}\}} \mathrm{Var}_{\frac{\ket{0}+\ket{N}}{\sqrt{2}}}(\hat n)& = \frac{N_{\rm max}^2}{4} 
    \nonumber \\
    &\implies \mathcal{F}_\theta= N_{\rm max}^2.
\end{align}
Since $N_{\rm RMS}\simeq N_{\rm max}$ along this path, this gives $\mathcal{F}_\theta\simeq N_{\rm RMS}^2$. We use this state only as a benchmark and do not analyze its preparation within $\mathcal{G}$.

Similarly, consider the displaced cat state $\ket{C}=(\ket{0}+\ket{\alpha})/\sqrt{2(1+e^{-A/2})}$, with $A=|\alpha|^2$, whose normalization includes the overlap $\braket{0|\alpha}=e^{-A/2}$. Its first two moments are
\begin{align}
    \langle\hat n\rangle_C=\frac{A}{2(1+e^{-A/2})},\qquad
    \langle\hat n^2\rangle_C=\frac{A^2+A}{2(1+e^{-A/2})},
\end{align}
so that
\begin{align}
    \mathcal{F}_\theta=4\,\mathrm{Var}_C(\hat n)=A^2+2A+\mathcal{O}\big(A^2e^{-A/2}\big).
\end{align}
For $n\geq1$, its photon-number probabilities are $q_n=p_n(A)/[2(1+e^{-A/2})]\leq p_n(A)$, where $p_n(A)=e^{-A}A^n/n!$ are those of the coherent state $\ket{\alpha}$. The cat state therefore satisfies the envelope whenever $\ket{\alpha}$ does. Taking $|\alpha|=|\alpha_{\rm tot}|$, the coherent state belongs to the Gaussian class, and Eq.~\eqref{Nmax_certified} gives, for large displacement, $N_{\rm max}\simeq e^{2r_{\rm tot}}|\alpha_{\rm tot}|^2$ for $|\alpha_{\rm tot}|^2\gg\ln\varepsilon^{-1}$. Hence $\mathcal{F}_\theta\simeq e^{-4r_{\rm tot}}N_{\rm max}^2\simeq e^{-4r_{\rm tot}}N_{\rm RMS}^2$, which reduces to $\mathcal{F}_\theta\simeq N_{\rm RMS}^2$ when the setup supplies no squeezing.

Finally, the QFI of the squeezed vacuum is $\mathcal{F}_\theta=8\langle n\rangle(\langle n\rangle+1)$. Along the path $\alpha_{\rm tot}=0$ with increasing $r_{\rm tot}$, $s_*$ approaches $\kappa\simeq2e^{-2r_{\rm tot}}$, and Eq.~\eqref{Nmax_certified} gives $N_{\rm max}\simeq e^{2r_{\rm tot}}(\ln\varepsilon^{-1}+2r_{\rm tot})/2$, up to corrections logarithmic in $\ln\varepsilon^{-1}+2r_{\rm tot}$. Hence $\langle n\rangle\simeq N_{\rm max}/[2(\ln\varepsilon^{-1}+2r_{\rm tot})]$, and
\begin{align}
    \mathcal{F}_\theta\simeq\frac{2N_{\rm RMS}^2}{(\ln\varepsilon^{-1}+2r_{\rm tot})^2}.
\end{align}
For $\varepsilon=10^{-6}$ and $r_{\rm tot}=2$, the exact minimization gives $\mathcal{F}_\theta/N_{\rm RMS}^2\approx5.3\times10^{-3}$.

\end{document}